# Pricing, liquidity and the
## control of dynamic systems in finance and economics

**Geoff Willis**

gwillis@econodynamics.org

## 0.0 Abstract

The paper discusses various practical consequences of treating economics and finance as an inherently dynamic and chaotic system. On the theoretical side this looks at the general applicability of the market-making pricing approach to economics in general. The paper also discuses the consequences of the endogenous creation of liquidity and the role of liquidity as a state variable. On the practical side, proposals are made for reducing chaotic behaviour in both housing markets and stock markets.

## Contents

# 1.0 Introduction

*"But there is one feature in particular which deserves our attention. It might have been supposed that competition between expert professionals, possessing judgment and knowledge beyond that of the average private investor, would correct the vagaries of the ignorant individual left to himself. It happens, however, that the energies and skill of the professional investor and speculator are mainly occupied otherwise. For most of these persons are, in fact, largely concerned, not with making superior long-term forecasts of the probable yield of an investment over its whole life, but with foreseeing changes in the conventional basis of valuation a short time ahead of the general public. They are concerned, not with what an investment is really worth to a man who buys it "for keeps", but with what the market will value it at, under the influence of mass psychology, three months or a year hence. Moreover, this behaviour is not the outcome of a wrong-headed propensity. It is an inevitable result of an investment market organised along the lines described. For it is not sensible to pay 25 for an investment of which you believe the prospective yield to justify a value of 30, if you also believe that the market will value it at 20 three months hence.*

*Thus the professional investor is forced to concern himself with the anticipation of impending changes, in the news or in the atmosphere, of the kind by which experience shows that the mass psychology of the market is most influenced. This is the inevitable result of investment markets organised with a view to so-called "liquidity". Of the maxims of orthodox finance none, surely, is more anti-social than the fetish of liquidity, the doctrine that it is a positive virtue on the part of investment institutions to concentrate their resources upon the holding of "liquid" securities. It forgets that there is no such thing as liquidity of investment for the community as a whole. The social object of skilled investment should be to defeat the dark forces of time and ignorance which envelop our future. The actual, private object of the most skilled investment to-day is "to beat the gun", as the Americans so well express it, to outwit the crowd, and to pass the bad, or depreciating, half-crown to the other fellow.*

*This battle of wits to anticipate the basis of conventional valuation a few months hence, rather than the prospective yield of an investment over a long term of years, does not even require gulls amongst the public to feed the maws of the professional; - it can be played by professionals amongst themselves. Nor is it necessary that anyone should keep his simple faith in the conventional basis of valuation having any genuine long-term validity. For it is, so to speak, a game of Snap, of Old Maid, of Musical Chairs - a pastime in which he is victor who says* Snap *neither too soon nor too late, who passes the Old Maid to his neighbour before the game is over, who secures a chair for himself when the music stops. These games can be played with zest and enjoyment, though all the players know that it is the Old Maid which is circulating, or that when the music stops some of the players will find themselves unseated."*

JM Keynes [Keynes 1936]

This paper is a much edited extract from a much larger paper 'Why Money Trickles Up' (henceforth YMTU) [Willis 2011]. In YMTU the author introduced a wide ranging modelling approach based on Lotka-Volterra and general Lotka-Volterra models. This approach proved capable of explaining the power laws seen in wealth, income and company size distributions. It also produced an economic model that explained the split between returns to labour and capital.

More interestingly for the discussions in this paper, the commodity and macroeconomic models in YMTU demonstrated that there could be substantial differences between the 'prices' of companies and commodities based on expected revenue streams, and the intrinsic or fundamental 'values' defined by input costs. The models demonstrated that this variation was inherent in the basic pricing system of capitalism as described by Minsky. The models are in fact simple and robust mathematical models of Minsky's qualitative descriptions.

This paper does not however require belief in the modelling of YMTU, it simply requires a disbelief in the efficient market hypothesis. This paper simply requires that the reader believes economies and markets are inherently dynamic, chaotic systems; that prices can become detached from fundamentals, that momentum chasing exists and that investors frequently estimate future results simply from historical trends.

The discussions in this paper are primarily the thoughts of a hands-on control engineer when faced with a dynamic, chaotic system. It consists of three parts. In section 2.0 there is a discussion of pricing in finance and economics and the parallels between market microstructure and post-Keynesian pricing theory. In section 3.0 there is a review of recent research in liquidity and a discussion of the consequences of liquidity as a state variable in economics. In sections 6.0, 7.0 and 9.0 some practical examples are given of how dynamic systems in economics and finance might be controlled. The first example in section 6.0 of controlling the housing market includes practical and sensible control measures. The examples in sections 7.0 and 9.0 look at stock markets and are far more speculative. It is however hoped that these examples may stimulate further debate on the effective control of chaotic markets.

(Nb the numbering of sections and figures in this paper partly follows the original numbering of 'Why Money Trickles Up'.)

## 2.0 Pricing in Finance and Economics

*An interesting puzzle in the history of economic thought is why the mathematization of economic theory in the 1940s and 50s took place through the formalization of the static Walrasian model, rather than through the study of infinite horizon production based models that arise from the Classical view. This puzzle is particularly intriguing because the best mathematician who ever worked on economic problems, John von Neumann, introduced the key mathematical tools in a study of such a Classical, infinite-horizon, production-based model before Arrow and Debreu used the same tools (mostly topological) to formalize Walras.*
[Foley 1990]

The idea that production rather than exchange is the source of value is contentious within mainstream economics, though why this is so is puzzling. Both the theoretical history and empirical data support this central view of production.

The theoretical debate goes back at least to the work of Sraffa and the Cambridge Capital Controversies. Sraffa's work demonstrated that the production function approach of marginalism was not appropriate, and that pricing of produced commodities through the long period classical approach was the appropriate way forward. The original work of von Neumann was also classically based, and also showed that a coherent system of prices can be built using the approaches of classical economics.

With regard to the clash between classical and neo-classical approaches, the work of Burgstaller [Burgstaller 1994] is particularly intriguing, in that he proposes that both the neoclassical and classical approaches can be presented as subsets of a unified approach.

In particular he shows that the neo-classical approach is appropriate when no labour is involved, as for example in a pure exchange process, while the introduction of labour results in the necessity of a classical approach. Burgstaller's work suggests that the neo-classical approach is only suitable for processes such as the purchase of raw materials and land, or interestingly, in the exchange of financial products.

In this light, marginalism would at first appear to be very useful in defining the mechanics of the purchase and sale of financial assets. With financial assets, owners have strong preferences for ownership, based on different preferences for risk, liquidity, etc. At a particular point in time, they will also have set initial endowments.

Following an exogenous event, such as an unexpected change in dividends, interest rates, etc, market participants will presumably want to rebalance their endowments to bring them into line with their preferences.

However, the financial field of market microstructure, with its wealth of data, has long moved on from simple static supply and demand curves.

Research in market microstructure has shown that the determinants of prices are stocks (inventories not shares), information, liquidity, etc, while marginality has been quietly sidelined. This is primarily caused by the problems of matching supply and demand over a time basis. When time is taken into account, marginality is replaced with a focus on inventories of financial assets owned, and the information encoded in order flow.

These conclusions on sources of costs are based on substantial quantitative research, supported by some very interesting theoretical work. This work is well reviewed in papers by Stoll, Madhavan and Biais et al, Stoll is a particularly good introduction [Stoll 2003, Madhavan 2000, Biais et al 2005].

Lyons also discusses this with great clarity in 'The Microstructure Approach to Exchange Rates' [Lyons 2001]. In sections 6.3 to 6.5 Lyons captures the difference between the 'Tastes & Technology' approach of traditional economics and the 'Information & Institutions' approach of market-microstructure. The utility approach of 'tastes & technology' rests on hypothetical foundations supposed in the late 19th century. The 'information & institutions' foundations are based on theoretical models proposed to fit large scale data sets through the finish of the 20th century and the start of the 21st. As Lyons notes: *"The microstructure approach also includes utility maximization, but as we saw in chapter 4, utility is specified very simply, typically in term of terminal nominal wealth."*

Market microstructure has analysed two main forms of markets, those composed of continuous double auctions and those made by market makers. With regard to pricing in general, the second is of particular interest.

Market makers buy and sell shares or other financial assets in financial markets. Financial markets involve buying and selling things in a dynamic time frame. There is no guarantee that somebody will want to buy something at exactly the same time that someone else will want to sell something. Market makers keep markets working by 'providing liquidity' and ensuring that there is always somebody who is willing to buy and sell shares at any particular time.

Market makers make markets by acting as intermediaries and do not normally hold on to shares on a long-term basis. They make their living by maintaining a small margin between the prices at which they buy and sell. Market makers are normally obliged by market rules to post prices at all times, and are obliged to fulfil purchases and sales at their advertised prices. They normally have to do this while in competition with other market makers. The speed of trading means that markets never formally 'clear' and market makers are often working 'blind' with little information other than the recent trading history of themselves and their competitors, and the knowledge of the level or inventory of assets that they currently have on their books.

Market makers make money by having a margin between the prices at which they sell and buy, this is known as the 'bid-ask-spread' or simply spread.

Market microstructure empirical research, experiments and theory have left the models of supply and demand behind; primarily because there is no evidence to suggest that market makers use marginality in pricing, and significant evidence that other factors are used in their pricing strategies.

Research suggests that the bid-ask spread is made up of five main components, these are discussed briefly below, for a more detailed review see Stoll, Madhavan or Lyons.

The first type of cost is administrative or 'handling' costs and other overheads. These reflect the costs of renting offices, paying wages, running systems etc. For modern electronic share-dealing these costs are generally very small.

Another cost may be caused by non-competitive practices, such as industry standards on tick sizes or standardised bid-ask spreads.

A third source of cost is related to the cost of holding unwanted inventory. Market makers are like bookies at horse races. Bookies probably know the horses and jockeys far better than the punters, but they don't make their money by betting on the horses. They make their money be balancing the supply and demand of the various betters, and making sure they take a small margin in the middle. It is dangerous for them to take a lot of bets on one horse, even if they think the horse will probably lose, because if it does win they will be wiped out. If they do get a lot of bets on one horse, even if they think the horse is lame, they will increase the price of that horse (by decreasing its odds) and decrease the price of the other horses (by increasing the odds) until they bring their positions back in to line and ensure that they will make a small profit whichever horse wins. In the same way market makers also generally know their markets much better than their customers. But they do not normally wish to hold large positions in a single stock, because if the price of that stock should collapse unexpectedly then they could go bankrupt overnight. Because of this managing and hedging inventory can be a significant source of costs.

This leads on naturally to the fourth source of cost, the cost of 'adverse information'. However well the market-maker knows his markets, he will never know them as well as 'informed traders', that is people who are closely linked or even working for the company whose shares are being traded, and so will have knowledge of good or bad news about the company before the market-maker. These 'informed traders' are able to make money out of the market-maker, and for the market-makers to stay in business, they must collectively recoup this money from the 'uninformed traders', they do this by having an appropriate extra margin in their bid-ask spreads.

A final source of costs is what is known as the 'free option' cost. In a well administered market, providers of liquidity are forced to hold their quotes open for a fixed minimum period. Priority rules then ensure that orders are closed out in a fair manner normally based firstly on price priority, then on time priority, where prices are equal. These rules force market-makers to compete with each other and so protect the ordinary share-trading public.

One problem with this is that it forces the market-maker to hold his price for a fixed time period; in this time the market price may move, giving an advantage to a well informed customer who can make money out of this 'free option'. To protect themselves, market makers add a small extra margin into the bid-ask spread.

As well as the work of pioneers in finance and economics covered in these papers, this area has also recently been extensively researched by others from the field of econophysics such as Farmer et al, Wyart et al and Bouchaud et al [Farmer et al 2005, Wyart et al 2008, Bouchaud et al 2009]. Taken together, the fields of market microstructure and econophysics seems close to providing full models for financial market functions that combine good theoretical underpinnings with good fits to actual data.

There also appears to be strong areas of similarity between the research that has been carried out in the area of market microstructure and that of post-Keynesian pricing theory. To the best of my knowledge these parallels do not appear to have been investigated.

Post-Keynesian pricing theory is primarily empirical, and its empirical basis is of a depth and surety rarely found in economics. In 'Post-Keynesian Price Theory' [Lee 1999] Frederic Lee gives an excellent review of how far disconnected from reality is the marginal approach to the pricing of manufactures. Despite the book's title, 80% of the book provides an excellent review of extensive historical research showing how businesses actually carry out pricing policy.

The results of the research show that, in the real world of business, marginality is non-existent. In particular, most businesses have their maximum profitability at maximum output. Diminishing returns simply don't appear in real world manufacturing. This has been clear for decades, see for example [Eiteman & Guthrie 1952]. In almost all production processes costs decrease with production right up to maximum output, and extra capacity, in the form of new factories, can be added easily and speedily. Under these conditions; of decreasing returns to scale, marginality is irrelevant as it simply cannot work.

In the real world almost all companies carry out their pricing using some variation of an average cost and 'mark-up' basis, with standard additional costs being added to the prices of the inputs. It is important to note that, as with market makers; manufacturers and retailers also price their goods in advance of sale when supplying to the public. They also often do so on long-term defined contracts when supplying to other companies. It is also notable from the post-Keynesian research that manufacturers and retailers focus strongly on inventory levels and the prices of their competitors for their decisions on prices and production quantities.

The existence of mark-up pricing and controls based on inventory levels, along with the absence of diminishing returns, is strongly supportive of the classical economists' point of view.

The parallels between post-Keynesian pricing theory and market microstructure theory are clear. Companies are obliged to behave as market makers. Complex market makers; but market makers none the less.

For a company, their 'mark-up' is directly analogous to the 'bid-ask spread' of the financial market-maker; though the weightings in the spread are a little different.

An easy example to follow is that of a retailer. A shop buys goods from manufacturers and sells the same goods on to the general public. So in this case the main inputs and outputs are identical; in the same manner as a financial market maker. While overheads for a financial market-maker are very small, for the retailer they are much larger, and need to pay for the remaining inputs of staff wages, distribution costs, rental of shop space, services, advertising, etc. They also need to include for payment of profit on capital and interest on debt. But just like stock markets, prices never formally 'clear', and pricing is based on information from competitors, rates of sales turnover, and levels of inventories of goods held.

Manufacturers, or providers of services, follow exactly the same logic, but now the stocks bought and the stocks sold are of different goods, and the 'bid-ask spread' is even larger and now includes the costs of the value adding processes used in production.

It appears that the substantial body of post-Keynesian empirical work could benefit strongly from looking at analytical ideas from market-microstructure and econophysics research.

Indeed the processes of market-making and market microstructure approaches in general appear to be ubiquitous and universally applicable in its role of price formation in economics as well as finance. Perry Mehrling provides a very thoughtful analysis of the US banking system using market microstructure approaches, while Lyons does the same for currency trading [Mehrling 2010, Lyons 2001].

The processes described by market microstructure concentrate on order flow and spread. They arise from markets in which prices are dynamic and not formally settled, where prices ultimately are linked to long-term values, but public information on those values is usually not complete.

In this price discovery process, information is found, and long-term prices are defined on different levels. Long-term prices will ultimately link to fundamental values, but as has been shown in YMTU, 'correct prices' will also vary with the point that has been reached in different cycles, on levels of liquidity and debt in the economy, on levels of government activity in the markets, on relative levels of trade and capital flows between different countries and levels of inventories of financial assets in the portfolios of different investors. As described in section 8.2.2 of YMTU, in such complex systems, the link to 'fundamental' values is weak and time dependent.

Market microstructure describes the mechanisms that allow buyers and sellers to discover these 'correct' values. Moving to a focus on inventory levels means moving to a world of dynamic equilibrium, of Lotka-Volterra models, predators and prey and maximum entropy production. The work of Sraffa, von Neumann, Kurz & Salvadori, Burgstaller, etc give a very good starting point for the calculation of long term prices in such a world. Unfortunately the approach used by these authors remains one based on static processes and single period analyses. Recasting this work into a dynamic approach should be straightforward. A sensible way forward would seem to be by using the market microstructure, market maker / post-Keynesian approach to attack the single-commodity, multiple-commodity, joint-production, etc, problems. If a simulation approach was used, rather than an algebraic approach, this might also reduce the ratio of headaches to results.

One of the main reasons for market 'prices' moving away from fundamental 'values' is the role of liquidity. This is discussed in section 3.0.

## 3.0 Liquidity

Liquidity is a measure of how easy or difficult it is to buy and sell things. It has been shown in the macroeconomic models of YMTU that liquidity can be artificially generated in a financial system simply by the known short-termism of markets combined with standard financial pricing procedures.

Liquidity has been the subject of much interesting research in recent years. This research suggests that liquidity could be of key importance in the apparent failure of markets to price assets correctly, and in the failure of financial markets in general. It does not appear that this research has so far made much impact in the fields of economics, finance or, with rare exceptions, in econophysics, which I believe is unfortunate.

The following is a brief review of current research and emerging ideas within the field of liquidity. The discussion is largely confined to liquidity within stock markets and its effects on the pricing and trading of stocks and shares.

Both the definition and measurement of liquidity presents problems. Historically stock market liquidity has been defined as the ability to trade large quantities of shares quickly, at low cost and with minimal price impact. Unfortunately this actually describes a range of desirable outcomes rather than an underlying concept or property.

Similarly, measurements of liquidity may focus on trading quantity, trading speed, trading cost, volume of trade, etc. Historically it has not been clear whether these different measures were in fact measuring the same thing or not.

In the last decade a number of papers have been produced giving comparisons of measurements of liquidity and illiquidity, see for example: [Chordia et al 2000, Porter 2008,

Korajczyk & Sadka 2006, Goyenko et al 2009]. Many different variables have been used to measure liquidity including trading volume, frequency of shares traded, bid-ask spreads, order imbalances, amongst many, many others. The variety of measures used reflects the difficulty of pinning down exactly what liquidity is. As well as individual measures, composite measures have been created in an attempt to capture the multiple dimensions of liquidity. Indeed there seems to be something of a cottage industry in the creation of new measures of liquidity.

In the more recent papers such as those above, it appears that more sophisticated measures of the different dimensions of liquidity do in fact correlate closely. It also appears that annual and monthly, long time scale data, correlates well with daily data [Goyenko et al 2009]. These results appear to hold true for both stock markets as a whole and individual company shares.

The research above suggests that the different measures of liquidity are in fact measuring the same underlying property, however the exact definition of this underlying property remains elusive.

It appears that including liquidity risk as a factor may explain a number of prominent 'market failures'. The following are given as examples:

Historically, domestic closed end funds have traded at a discount to the underlying shares, while international closed end funds have traded at a premium. These results can be explained by the greater liquidity of the domestic shares vis-à-vis the funds, and the less liquid foreign stock markets compared to the US fund share market [Amihud et al 2005 – 3.4.5].

Similarly, in most countries, where companies have two classes of shares for nationals and foreigners, the national owned shares trade at a lower price than foreign owned shares. In China the reverse is true. This appears to be a consequence of the high level of liquidity in the Chinese domestic stock market [Chen & Swan 2008], while in most countries the domestic market is less liquid than international markets.

Similar arguments can be used to explain the discounts on restricted stocks [Amihud et al 2005] as well as the differences between prices of treasury notes and treasury bills [Amihud et al 2005 – 3.3.1] and also of treasury notes versus corporate bonds; where the price difference can not be accounted for by default risk alone [Amihud et al 2005 – 3.3.2].

Chordia et al, have demonstrated that liquidity problems can explain the post earnings drift that follows unexpectedly high or low earnings announcements [Chordia et al 2009]. While Korajczyk and Sadka show that liquidity can explain up to half the benefits of momentum strategy anomalies documented by Jegadeesh & Titman [Korajczyk & Sadka 2006].

To date I haven't seen a paper discussing the anomalies of dual listed companies such as Royal Dutch Shell, however I confidently expect liquidity to explain the long-term diversion of such share prices.

While all the above are interesting, probably the most important result of recent research into liquidity, is that liquidity, or more correctly, liquidity risk appears to be a major component of asset pricing.

Amihud et al, give a full review of these results, which demonstrate that a liquidity augmented Capital Asset Pricing Model (CAPM) gives much better results than a traditional CAPM [Amihud et al 2005 – 3.2.3]. Other work supporting this view has been carried out by Acharya & Pederson and Pastor & Stambaugh using single measures of liquidity [Acharya & Pedersen 2005, Pastor & Stambaugh 2003], Goyenko et al, Korajczyk and Sadka [Goyenko et al 2009], Liu [Liu 2006 & 2009] and Lee [Lee 2005].

Given the poor historical performance of the CAPM, the Fama-French three factor model has often been used as an alternative. This uses firm size and book-to-market ratio in addition to a market index. Results from the research above strongly suggest that a single liquidity measure can replace both firm size and book to market ratio and give improved results. This suggests that both firm size and book to market ratio may be surrogate measures for liquidity risk.

Fama and French's own work indicated that as well as the factors of risk, firm size and book to market ratio, a fourth momentum factor needs to be included to fully explain share price movements [Fama & French 1992]. If the research in liquidity stands up to further investigation, it suggests that share price movements can be explained by just risk, liquidity and momentum.

Further to that, and in line with the workings of the macroeconomic models of section 4 of YMTU, the work of Korajczyk and Sadka [Korajczyk & Sadka 2005] suggests that provision of liquidity also reinforces momentum strategies. This suggests that short term momentum pricing is not 'behavioural' or even plain stupidity, but is 'rational' behaviour for participants, until the market finally reaches a position far out of equilibrium, and endogenous liquidity creation is stalled.

Taken together it appears that a new 'three factor' asset pricing model involving the market beta, liquidity risk, and momentum may be superior to both the CAPM and the Fama-French 'three factor' model.

This then becomes much more significant at the level of the whole stockmarket, especially in the light of the extensive work by Shiller and Smithers regarding the long-term valuation of stock markets. This work is very well summed up in 'Wall Street Revalued' [Smithers 2009].

The central thesis of this work is straightforward. Shiller and Smithers find that stock market prices do not follow random walks, but are in fact mean reverting over decadal timescales. Two measures in particular are able to capture the over or under valuation of the stock market, the two measures that do this are CAPE and Tobin's q.

Tobin's q is of course the same thing as the book-to-market ratio, the same value used at company level in the research of French & Fama and various other researchers in liquidity. q is just the ratio of K to W in the models of YMTU.

At a whole stock market level, both company risk factors and company size are averaged out, leaving only book to market value as a meaningful indicator.

It appears that by measuring the value of Tobin's q, researchers such as Shiller and Smithers have simply been measuring the liquidity of the whole stock market, with Tobin's q acting as a close proxy measure for liquidity.

On the other hand the 'CAPE' is the 'cyclically adjusted price to earnings ratio', which is simply the price to earnings ratio adjusted to a long time period; normally ten years. The CAPE also provides a very good measure of over/under valuation, and consequently correlates very closely with Tobin's q.

Working backwards, the logical conclusion is that the over- or under-valuation of the stock market, defined by long term earnings and prices, is simply a measure of the overall liquidity in the stock market, and that deviations away from the long term average are almost wholly due to liquidity.

The anecdotal evidence that equity prices are linked to liquidity is certainly plausible. The dramatic fall in share prices during the 2008 Credit Crunch and the subsequent rebound following the introduction of quantitative easing and other fiscal loosening are strongly suggestive of a direct link between liquidity in the economy as a whole and equity prices.

To date there appears to have been relatively little research in this area, which is unfortunate considering its potential importance.

Pepper & Oliver [Pepper & Oliver 2006] have produced an extensive study of this issue. Their work is very persuasive, and an excellent discussion of how liquidity works in practice, but the attempts to link share price levels to monetary data, while compelling, are not conclusive. This reflects the problems of finding trustworthy monetary data, a problem that the new approach using liquidity measures may alleviate.

More recently, Chordia et al, Jones, and Pastor & Stambaugh, have used different measures of market liquidity and have all noted correlations of liquidity to market movements. [Chordia et al 2001a, Jones 2002, Pastor & Stambaugh 2003].

Liu has carried out a longer and more detailed analysis and concludes that there is evidence for changes in liquidity corresponding to market movements, and that this is consistent with the argument that liquidity is a state variable important for asset pricing [Liu 2006, 2009].

Chordia et al have carried out an empirical analysis of the relations between liquidity in the stock, bond and money markets, and suggest important links between liquidity, volatility, and monetary policy [Chordia et al 2005].

While it is early days, it appears that not only is liquidity of fundamental importance in the pricing of stocks and other financial assets, it appears that it may in fact be a fundamental state variable of financial markets, and one that is straightforward to measure on a timely basis. If this is true, then there are some big implications for both finance and economics.

Historically, attempts to measure liquidity at a national level have focused on measurements of money supply. Most notably in the UK in the early 1980's monetary policy was used in an attempt to control the economy. The policy was quickly discredited, primarily due to the difficulties of collecting timely and accurate monetary data, and also due to the ease with which the sources of such data could be manipulated by financial institutions, see Pepper & Oliver for more details.

In marked contrast, some of the liquidity measures used in more recent liquidity research, for example those of [Chordia et al, 2002 & 2005], are easily calculated on a daily basis from stock market information. It would be trivially easy for indices and sub-indices of liquidity to be set up that could be observed and used by both the financial markets and economic actors.

The research on liquidity suggests two implications for finance that are both quite profound, the first area relates to the pricing models based on Black-Scholes, the second to the pricing of shares under the CAPM.

Almost all modern option pricing theory is based on the Black-Scholes model, or other closely-related models. Black-Scholes has been one of the most important mathematical contributions to economics or finance, and certainly the only one to have come into widespread day to day use within the financial industry.

However, one of the core assumptions of B-S is that options on shares, as well as the underlying shares themselves, can be bought and sold easily in highly liquid markets.

The recent body of work studying liquidity of financial assets suggests that this assumption is profoundly flawed. It seems likely that prices of both options and underlying assets will be affected significantly by liquidity. It also seems likely that the effects might not be the same for the option and the underlying. Consequently this would suggest that B-S models would, as a minimum, need modifying to take into account the effects of liquidity.

That liquidity should be a concern for quantitative finance in general seems obvious; Long Term Capital Management (LTCM) was brought to earth largely through trading in products that became illiquid overnight, and illiquidity was a major factor in the collapse in asset prices that took place during the credit crunch.

Clearly the effect of liquidity on asset prices appears to be an area ripe for more quantitative analysis. The possibility of a relationship between liquidity and volatility seems particularly interesting. Other than the work of Chordia et al [Chordia et al 2005] discussed above, there appears to be little published research in this area.

If it is true that liquidity is an easily measurable state variable of shares, and that also there are mathematical relationships between liquidity and volatility (which seems plausible), then it may be that measurement of liquidity might be able to give good timely measures for current volatility that can be used directly in Black-Scholes models; rather than the current practice of imputing from historical volatility.

A second significant area of interest for the application of liquidity in finance is to asset pricing models. The research to date suggest that liquidity can replace both the size and book to value elements in the Fama-French three factor model, leaving only risk and liquidity, along with momentum, as the determinants of equity prices. Or to put it another way, liquidity risk appears to be the main missing risk element of the various CAPM models. This knowledge gives the intriguing possibility that it should be possible to fully hedge an asset portfolio, and, more questionably, that this might even lead to self-stabilising markets in asset prices.

As discussed above, some of the liquidity measures are easily calculated on a daily basis from stock market information.

It would be trivially easy to set up a standard 'liquidity index', similar to the VIX index for volatility, and encourage trading of futures in the index and so allow a deep market to form in this liquidity index.

Investors would then be able to go long on shares, or stock market indices, and simultaneously short the liquidity index to protect against a reduction or collapse in liquidity. If the recent work on liquidity is correct, this should give almost full protection on an asset portfolio of investments.

Interestingly, this should act in a strongly counter-cyclical manner. Given the mean reversion properties of the market as per Shiller and Smithers, liquidity protection of this type should be cheap at historical liquidity lows, but increasingly expensive as liquidity bubbles formed; if, of course, it was correctly priced.

If such hedging functioned correctly, the cost of protecting against excessive liquidity would itself prevent excessive overpricing of assets and would automatically withdraw liquidity from the market as prices became frothy.

As well as having an overall liquidity index, there would also be scope for sub-indices tracking individual sectors. Indeed it may make sense to re-sort companies from traditional 'industry' sectors into groupings that share a common pattern of historical liquidity and volatility behaviour.

Clearly correct pricing, and the formation of a sufficiently deep market to cover even a portion of the stocks traded might be problematic. There are also clear possibilities of counter-party default dangers of the sort that afflicted AIG following their substantial underpricing of CDS risk.

If liquidity risk is the main missing factor in the CAPM model, and also it proves possible to enumerate and hedge against this risk, then by analysing the resultant data, it may be also be possible to analyse and quantify the remaining residual risks in the pricing of assets.

In an ideal world, under these circumstances, it seems possible that momentum trading would become difficult and short-term speculation might be a difficult and profitless activity. This could lead to financial investment becoming a predictable and rather dull area of both business and economics. Common sense, and the weight of history, does suggest that this is more likely to be a possibility rather than a probability.

However, if deep and efficient markets in liquidity futures did form, then speculative interest would allow liquidity index pricing to change in response to external factors such as government policy, oil shocks and other exogenous events. This leads to the possibility that liquidity measures could also be very useful for macroeconomic control.

Having liquidity indices of this form could assist governments in targeting liquidity in stock markets, and in the economy in general. This might answer the problem of the poor quality and timeliness of traditional monetary data.

Casual observation suggests that there is poor short-term correlation between the supply of liquidity to financial markets and the health of the economy as a whole. In the United States for example, in 2005 and 2006 the stock market was booming, with very high liquidity, even though the economy as a whole was struggling (see also discussion of the Bowley squared model of section 4.9 of YMTU).

In such circumstances, central bankers face acute problems. With the single tool of interest rates, governments are in a cleft stick. This was admitted to recently by Kate Barker, an ex-member of the UK monetary policy committee [Guardian 2010]. In 2005 the UK appeared to be in both a housing bubble and a stock market bubble, but the general economy was sluggish, and inflation was historically very low, with the threat of deflation in the wings. Raising interest rates to calm down the housing and financial markets risked initiating a recession, possibly moving into outright deflation. However, failing to raise interest rates caused an ongoing bubble to continue its expansion, which had very unfortunate consequences including the collapse of Northern Rock, and the bailing out of Bradford & Bingley, Royal Bank of Scotland and other institutions.

As discussed above, in the past, attempts have been made to control the economy through the control of the money supply. Historically these attempts have not worked well, partly because the money supply is difficult to quantify and measure reliably.

I believe a second problem is that there are two prime sources of liquidity. The money supply and debt is one of them, but the endogenous creation of liquidity within the pricing system, described in YMTU, is another, and in my view this is the prime source and the larger source. So, certainly increasing the money supply and debt can increase liquidity in the stock market. But increases in stock valuations also create their own liquidity, and also provide apparent extra wealth against which new debt can be secured. These two sources of extra liquidity feed on each other in a most unhealthy way.

I believe targeting a liquidity measure in stock markets may be more effective than monetary targeting, as a liquidity measure is measuring the output, the residual, of the liquidity creation process. A certain amount of debt and new money supply is needed in an economy. If insufficient is supplied, then the stockmarket declines, if too much is provided the stockmarket booms, the stockmarket is normally a good weather vane for liquidity in the economy as a whole.

The macroeconomic models in section 4 of YMTU suggest that liquidity can be formed endogenously, in exactly the way proposed by Minsky. This suggests that, just as central banks are expected to control changes in the money supply caused by fractional reserve banking, it seems appropriate that they also be obliged to control money supply growth caused by Minskian asset price bubbles.

The recent research in liquidity, and the models of YMTU, suggest that liquidity needs to be targeted separately, in addition to the inflation targeting of the overall economy. The ease and timeliness with which liquidity can be calculated, and compared to historical liquidity levels, suggests that this would be relatively straightforward to do.

For instance it might be possible to use active management of the bond market as has recently been done under 'quantitative easing', on a regular basis to increase and decrease the liquidity of financial markets generally. So in 2005-06 it might have been sensible to actively embark on 'quantitative tightening' to restrain the financial markets, while simultaneously lowering interest rates to assist the larger non-financial economy.

An important caveat here is the role of housing, which appears to be more important than even the stockmarket as a driver of booms and busts. Controlling liquidity and money supply for an economy will only be effective if the housing market is stabilised. Absent an effective measure of liquidity in the housing market, then other damping measures and long term indicators need to be used such as historical ratios of house prices to wages and ratios of mortgage payments to rents.

**6.0 Dynamic Control of Housing Markets**

Figure 6.3.1 here

Figure 6.3.1 above shows the prices of housing in the UK from 1953 to 2010, divided by the average wage, prepared using data from the Nationwide Building Society and the UK Office of National Statistics. The high house prices immediately following the Second World War were a consequence of substantial loss of housing during the war and a suspension of house construction for the six-year duration of the war.

During the 1950s and 60s access to mortgages in the UK was tightly regulated and controlled by government micro-management of financial institutions, with direct lending ceilings imposed on banks and building societies; resulting in strict rules on eligibility, deposit sizes, etc.

During this period house prices showed remarkable stability at a cost of roughly 3.0 to 3.5 times average salary. It is very important to note that, despite the strong state controls on access to housing finance, the 50's and 60's were a time of substantial private house building in the UK. Despite the restrictions imposed by the state, even at these regulated 'low' prices, demand created lots of supply. As can be seen in figure 6.3.2 below UK private house building reached a prolonged peak in the mid 1960s.

Figure 6.3.2
[ONS 2004]

Access to mortgages was liberalised in 1971 this resulted in the 'Barber boom', clear in figure 6.3.1, stimulated by the resulting rise in liquidity. From the 1970's onwards, the UK housing market has been characterised by vicious cyclic booms and busts, with a very clear reversion to the pre-Barber long-term trend at, 3 to 3.5, at the bottoms of the cycles.

These cycles are identical in form to the ones discussed in the commodity models of YMTU. It is important to note that at the bottom of both the actual housing data, and the commodities models, prices reach their 'real', 'fundamental', Sraffian values. At these prices the value of housing represents the cost of the inputs. The same can also be seen clearly in data from the United States (this time deflated for cpi); see figure 6.3.3 below.

Figure 6.3.3 here
[Shiller 2010]

The persistence of these cycles is deep within the economy of the UK. In his book 'Boom, Bust, House Prices, Banking and the Depression of 2010' [Harrison 2005] shows that the cycles in the UK go back to at least the middle of the eighteenth century.

As an economic experiment, you could scarcely ask for clearer data output. The basic system dynamics are substantially and dramatically changed following a point change in policy in 1971. Not only that, but this experiment has controls; Germany and Switzerland for example, have retained strict controls on mortgages for house purchases and don't suffer from strong cyclical booms and busts in house prices. Figure 6.3.4 below has the average value of house prices included for the two periods.

Figure 6.3.4 here

The net result of the liberalisation of credit in 1971 was the increase in average cost of housing for all Britons by roughly 23%. In the last cycle, from 1996 to 2010, prices were fully 40% higher than the '55-'70 baseline rate.

Housing suffers from the same problem as capital-intensive commodities, as modelled in section 3 of YMTU. Construction of housing takes a finite time, and so house prices can go up significantly before market mechanisms have time to work. Unfortunately, housing also has the same problems of endogenous liquidity creation that is seen in the macroeconomic model of YMTU. As house prices go up, people feel richer, and also as with shares 'momentum' kicks in, and house prices, and the economy as a whole keeps rising, until finally house prices become unaffordable for new entrants in the market, and the bubble bursts. As a capital-intensive industry, housing is naturally cyclical.

This again shows that the contrast between the comparative statics of neoclassical economics, and the real world of dynamic differential equations is stark. With comparative statics it is easy to 'prove' that credit controls and other government interventions 'must' increase the price of goods, and so reduce the welfare of the public. So neoclassical economists always push for removal of such controls.

In the real world, where speculative cycles can be endogenously created within the economic system; credit controls and other 'interferences' in the market work beneficially by 'damping' the cyclical behaviour. It may be counter-intuitive, but in the right circumstances, applying controls and apparent 'costs' to the market actually reduces the price of goods. And reduces them substantially. In the area of UK housing, the data above shows that the reduction in prices would be over 20% if strict credit controls were reimposed tomorrow as they were in the '50s and '60s.

Going back to figure 6.3.1 or 6.3.3 for the UK and US it is clear that the 'investment' value of housing is a chimera. Over the long term, growth in the value of houses is derisory and barely keeps up with the growth in earnings. Stock market growth is typically 5% higher than this.

Smithers discusses the dual properties of housing as both a form of consumption and investment in Wall Street Revalued p 107-108 [Smithers 2009]. The fact that housing is fundamentally consumption is demonstrated by the continued reversion to a fixed proportion of wages.

Figures 6.3.1 and 6.3.3 show clearly that in the long-term housing is a proportion of wages, and behaves as consumption. Governments should treat it as such, and actively prevent houses being treated as investments, and most certainly should prevent them being treated as speculative investments.

Looking both at the UK data and the US data in figures 6.3.1 and 6.3.3, a very worrying development is that in both countries the size of the booms is steadily rising, though the falls back to normal are the same. From a controls point of view this is very worrying, it suggests that the cycles could be even more dramatic and dangerous in the future.

Faced with a dynamic, cyclical system, standard control systems knowledge can be used to control the system. There are two ways to remove cycling (what engineers call 'hunting') in a control system.

One is to use deliberate counter-cyclical feedback; most central banks try to do this using interest rates to control the economy as a whole. As central bankers are only too aware, this is not an easy way to control anything. A good example of such a feedback loop is a domestic shower system. A combination of a difficult to use mixer valve, and the delay between making the change at the tap and feeling the change in the water temperature often results in alternating flows of water that is too hot or too cold .

Wherever possible, a much better solution is to use damping of the cycle. When done successfully this can result in a dramatic drop in oscillations with fairly minor, adjustments to the system. This is like the example of using shock absorbers with a car's wheels to prevent the car vibrating wildly on its springs every time it hits a bump.

The strict credit controls used in the UK prior to 1971 provided just such an effective damping system. If all else fails it is imperative that such controls are reintroduced in the UK.

However it may be possible that less draconian measures may be just as effective.

As a rule of thumb, to be effective, damping measures need to have a time span of a similar order to that of the natural cycle time of the system, as a minimum they should be of a length of half a cycle or so. For the UK Harrison [Harrison 2005] shows strong evidence for a fifteen to twenty year cycle for house prices. Sensibly, damping measures need to be of the order of ten years or so.

Looking closely at the US data in figure 6.3.3; there is the same flat trend as the UK at the bottoms of the cycles; showing the same reversion to real, non-speculative, prices. It is also clear that the booms are a relatively new phenomenon.

A subtly different experiment has been carried out in the US. The change in behaviour of the housing market appears to be correlated with the rise in non-standard mortgage products. Historically the US has used fixed-rate mortgages, only moving to adjustable rate mortgages comparatively recently. In the UK adjustable, or short term fixed mortgages have been the norm for many years, and it is very difficult to get fixed rate mortgages of more than five years.

The finance industry does not like fixed-rate mortgages. It leaves the issuers holding interest rate and inflation risk. Moving to adjustable rates gives the appearance of moving the risk to individual mortgage holders. This in itself is a practice to be questioned in a democratic society. Why sophisticated finance companies should be allowed to offload complex financial risk onto individuals with little mathematical, let alone financial, training is not clear.

In reality, offloading risk in systemic fashion like this simply creates systemic risk. As has been made abundantly clear in recent years; ultimately the only realistic holder of systemic risk is the taxpayer. Allowing financial companies to issue variable rate mortgages is to give the financial companies government subsidised one-way bets.

Figure 6.3.5 below gives a comparison of mortgage types issued in various different countries in Europe.

6.3.5 here

[Hess & Holzhausen 2008]

The mainly variable countries are Greece, Spain, Ireland, Luxembourg, Portugal, Finland and the UK. This pretty much speaks for itself.

The solution to this is trivially straightforward. All loans that are secured against domestic property should be limited to a ten-year minimum and a thirty year maximum. They should also be fixed rate, or, as a minimum, be a fixed percentage above rpi or cpi, throughout the period of the mortgage. This would move interest rate risk back on to the shoulders of the finance industry. Where it belongs. Variable rate mortgages should be strictly illegal in any self-respecting democracy.

There are other sensible mechanisms to reduce the use of houses as investments, especially as speculative investments. The most obvious one is to have a capital gains tax that is more punitive than that for other investments. The tax should be charged on all houses, including first homes, without exception. Sensibly this would be a tapered tax; starting at say 20% for the first year, then drop by two percentage points per year, so reaching zero after ten years of ownership.

A much better approach would be to have a sales tax on all houses. This should be applied to the seller of all houses, whether they have increased or decreased in value. Again, sensibly, the tax should be tapered over the years.

A tapered capital-gains tax or house sales tax, with a ten-year taper should bring in the damping of the sort required to deal with a 15 to 20 year endogenous property cycle. People buying houses to live in would not be punished, speculators would be.

In addition annual property taxes, or land taxes, should be charged on the value of houses or on the value of the underlying land, rather than on the occupants, as many local taxes are.

Another sensible policy would be to have compulsory mortgage indemnity guarantee (MIG). House purchasers would be obliged to take out insurance to cover full potential losses against potential negative equity, ie the difference between mortgage loan value and likely sale value of house. Such insurance would be cheap if the purchaser had a large deposit and prices were below the long-term trend. The insurance would be very expensive if the deposit was small and it was the height of a boom. As such, compulsory MIG should act in a strongly counter-cyclical manner.

Many countries enforce minimum deposit requirements [Hess & Holzhausen 2008]. This seems a very sensible policy, as those with small deposits are far more likely to default, see for example figure 6.3.6 below.

6.3.6 here

[FT/S&P 2010]

It can be seen that arrears rates increase dramatically as deposit sizes reduce. As with variable rate mortgages, when governments allow financial institutions to offer low deposit rates; that is highly leveraged asset purchases, they allow financial institutions to offload their risk onto the state.

There is a more sophisticated and better way of addressing this particular risk problem. Rather than prescribe laws on deposits, a more effective law would define a maximum limit of say 80% of the sale value of a house that could be repaid to pay off debt secured on the property.

So if a homeowner was foreclosed on, and their property was sold off, a minimum of 20% of the sale proceeds would go to the homeowner, and the other 80% would be shared by all the creditors who have loans secured on the property. This would have a number of advantages. It would have the same effect as a minimum deposit requirement of 20%. Banks would generally be reluctant to supply a mortgage of greater than 80% of the value of the house. It would also make it much more difficult to evade the minimum deposit rules by taking out secondary loans secured on the house.

More subtly it would also act in a counter-cyclical manner. When house prices were at historical lows, banks might be willing to lend 90% mortgages, confident that house price were likely to rise. Conversely, when house prices were significantly above their long-term averages banks would require larger and larger deposits due to their fears that house prices might drop in the future. Similarly they would be very reluctant to allow mortgage equity withdrawal.

In addition to the passive management techniques discussed above, there is also a strong case for active counter-cyclical monitoring and management of the economy by central banks and other monetary authorities. Despite protestations to the opposite, housing bubbles are very easy to spot.

The first obvious measure is that shown in figures 6.3.1 and 6.3.3 for the US and UK. The ratio of house prices to median wages shows very strong patterns of reversion to mean.

Similar patterns are also seen in ratios of housing costs to rental costs. When house prices are correctly valued, housing costs (mortgage payments, etc) are close to rents on equivalent properties [FT 2010].

If either of these ratios increases significantly above the long-term trend then you are moving into a housing bubble.

At this point the central bank should intervene to prick the bubble as early as possible. This could be by increasing the sales tax or capital gains tax on houses, increasing deposit and MIG requirements or by imposing a tax on mortgage debt.

Finally, if none of the above work effectively to damp markets then the necessary solution is to simply bring back the same credit controls that the UK had prior to 1971.

## 7.0 Low Frequency / Tobin Trading

*THE spectacular collapse of so many big financial firms during the crisis of 2008 has provided new evidence for the belief that stockmarket capitalism is dangerously short-termist.........  Shareholders can no longer with a straight face cite the efficient-market hypothesis as evidence that rising share prices are always evidence of better prospects, rather than of an unsustainable bubble.*

*If the stockmarket can get wildly out of whack in the short run, companies and investors that base their decisions solely on passing movements in share prices should not be surprised if they pay a penalty over the long term. But what can be done to encourage a longer-term perspective?......*

*In the early 1980s shares traded on the New York Stock Exchange changed hands every three years on average. Nowadays the average tenure is down to about ten months. That helps to explain the growing concern about short-termism. Last year a task force of doughty American investors (Warren Buffett, Felix Rohatyn and Pete Peterson, among others) convened by the Aspen Institute, a think-tank, published a report called "Overcoming Short-Termism". It advocated various measures to encourage investors to hold shares for longer, including withholding voting rights from new shareholders for a year.* [Economist 2010a]

Warren Buffet is of course a value investor, the sort of investor who intuitively understands the workings of the companies models in section 2 of YMTU. The sort of investor that the efficient market hypothesis states cannot exist. Value investors also intuitively understand that the short-term liquidity and momentum effects seen in the commodity and macroeconomic models in sections 3 and 4 of YMTU not only make value investing difficult, but also add no value to the process of creating wealth that capitalism aspires to.

The proposals of the Aspen Institute were pretty much stillborn for a number of reasons. Firstly, because orthodox economics assumes, erroneously, that any cost imposed on market transactions must increase costs to the consumer. Secondly because such a tax would destroy a substantial part of the finance industry, which makes the majority of its profits by charging rents on the very volatility they create in the first place. And thirdly, and more reasonably, if such a tax were imposed in one country, trading would simply move to an alternative jurisdiction.

To understand just how short-term the finance industry has become, it is worth noting that stock-trading is now dominated by 'high-frequency trading' (HFT). In the major stock-markets supercomputers trade billions of dollars of trades in seconds using automated algorithms. Individual bids and offers may be held open for fractions of a second. High frequency trading systems are now being co-located within stock-exchange buildings as the speed of light now means that companies trading from a few blocks away are at a significant disadvantage.

To anybody who has actually worked in a real company, the idea that the real market value of a normal company can change from millisecond to millisecond is bizarre.

It is my belief that Buffet, Shiller, Smithers et al are correct, and that the unnecessary volatility is induced endogenously in share markets, causing excessive movements away from real value on timescales from seconds to decades.

It is my belief that the decadal movements are caused by liquidity at a macroeconomic scale, a problem that will need tackling at a macroeconomic level – this was discussed above in section 3.0.

Other timescales are much shorter and give the appearance of being quasi-periodic momentum effects. Although the evidence is controversial, typical time-scales for the periodicity appear to be on the order of fifty and two hundred trading days, with other shorter time scales also present. A system is proposed below that would dampen the fluctuations on these timescales.

The solution proposed is a private-sector approach, independent of government. Following the same logic as housing in the previous section, it is proposed to introduce damping with losses imposed on early retrading on the lines of those proposed by Buffet et al. This would be done by introducing a new class of shares, or special investment certificates, in the companies. These shares would have different rules as to their trading. The issuing of such shares would be voluntary, at the choice of the companies involved.

In the same way as housing, damping would be imposed with a haircut of say 10% imposed on anybody who sold a share within the specified time period. The haircut would be paid back to the company in which the share is held at the time of sale, as such it would be effectively a 'negative dividend' on the share, paid by the owner to the company. The haircut would automatically be deducted from the sale proceeds. In extremis the haircut would be imposed for a period of say three years.

However unlike housing it is not proposed that the haircut on all shares be imposed for the full term of three years. This would present great problems for pricing of the shares. If a large purchase was made of a company's shares this would kill the market in that company's shares for years at a time, which would make price discovery for the company almost impossible.

Instead it is proposed that all shares that have been sold are marked as 'locked'. This would be in contrast to all the remaining shares that would be 'unlocked'.

Every trading day a random selection would be made across all the currently 'locked' shares and 1% of all the currently locked shares would be unlocked. The owners of these newly unlocked shares would then be able to sell the shares immediately without penalty.

Assuming 250 days of trading per year, then this release of 1% of shares per trading day would give a half-life for locked shares of roughly six months.

This means that if every single share was bought on day one, and no further trading took place, roughly half the shares would be unlocked after six months, more than 70% would be unlocked by the end of the first year, over 90% would be unlocked by the end of the second year and almost 98% would be unlocked by the end of year three. At this point, after three years, any remaining locked shares would be automatically unlocked.

This system would be a compromise between ensuring a haircut on fast resellers, while ensuring that shares were continually made available to the market for further trading. For an individual purchaser who bought a block purchase, their haircut on day one, if they resold all their shares would be 10%, if they sold all shares after a year the haircut would be slightly below 3%, after two years it would be 1%. After three years the haircut would be zero.

In these circumstances purchasing shares for value investment would have very little risk as in such a circumstance the period would be expected to be a minimum of a few years. Speculative investment would be risky, and effectively pointless.

Even better for value investors, it should be noted that the losses taken by the early sellers accrue to the company in which the shares are held, and so ultimately to the other shareholders. The losses of the speculators are transferred directly to the value investors.

All of this could be simply organised electronically through the same systems that currently manage dividend payments.

Interestingly, although such a system may seem complex, it may actually be one that would be driven to adoption by the market. For well managed companies, issuing such shares would give direct benefits to value investors, but much more importantly issuing such shares would in its own right be a very powerful signalling mechanism to the market. It would be very foolish for a company that is manipulating a short-term rise in its share price to issue such shares, the subsequent burning of locked-in investors would cause significant reputational loss. On the other hand, for well-run companies with long-term investment horizons, issuing such shares would be a way of signalling the long-term commitment of the management. This would particularly be the case if managers share options were restricted to these shares. Eventually, failing to issues such shares might become a good indication of a poorly managed company.

Such a shareholding pattern might form a useful compromise between the pattern of 'Anglo-Saxon' free trading of shares and the 'European' model of very long-term share-holding with very low levels of open trading.

# 9.0 Improving Continuous Double Auctions

In the previous section I was critical about the fashion for high-frequency trading. In an act of some foolishness I would like to look at this in more detail. I do this with some trepidation, moving into an area where debate is vociferous and my knowledge is limited. However, despite my inexperience, from my naïve viewpoint it appears that the structure of financial markets often seems perverse and appears to be incentivised against easy price discovery and the simple execution of large trades.

As discussed previously, stock trading is now dominated by 'high-frequency trading'. On the major western stock markets the majority of trading is done by high-frequency algorithmic trading. In these stock markets supercomputers trade billions of dollars of trades in seconds using automated algorithms. Individual bids and offers may be held open for fractions of a second.

This is done in the sacred name of 'liquidity', which is assumed to be always a good thing.

The current data suggests that high-frequency traders largely provide their liquidity to well-traded shares in preference to infrequently traded ones. They also prefer doing so at times of low volatility to high volatility. By definition this is opposite to the requirements of effective liquidity supply, and the reverse of a couple of centuries of defining the role of liquidity suppliers.

The quote from Keynes at the beginning of section 1.0 gives his views of the benefits of liquidity, and it appears reasonable to assume his opinion of high-frequency trading would not have been positive.

More recently other experienced financiers have shared similar views [Noser 2010], and at least one commodity trade body has denounced 'parasitic' traders [FT 2011b].

That my concerns are more widely held is supported by the recent decision of Credit Suisse to start a 'light-pool' for institutional investors. This is deliberately aimed at large volume traders and 'opportunistic traders' will be specifically denied access to the system [FT 2011a].

My own fundamental problems with high-frequency trading are three fold.

Firstly it is trivially obvious that the value of companies does not change from microsecond to microsecond. In fact research suggests that publicly announced information has negligible effect on trading, see for example [Joulin 2008, Ranaldo 2008, Bouchaud et al 2009]. In fact information largely comes from large trades by institutional traders, and as Bouchaud et al make clear, the savagery of the market means that such large trades now need to be broken up into small trades and fed into the markets in a piecemeal fashion, sometimes in periods as long as months, to prevent adverse price movements.

This brings me to my second fundamental problem with HFT. In a dynamic, chaotic system, reducing the time constant of trades, allowing trades to be faster and faster, increases the speed and volatility of short-term momentum processes.

To go to the idea of a traditional market, if I was a customer trying to buy or sell oranges from or to a stall-holder, I would naturally prefer to see all the stall-holders displaying their prices while I get the opportunity to walk around and chose the best price. If each stall holder just flashed a quote for one second and told me to take it or leave it, things would be much more difficult for me.

Finally, and leading on from the above, there is very little evidence that high-frequency trading does in fact provide liquidity. The paper by Bouchaud et al is magisterial in its depth, and the main conclusions are that, although a lot of shares are traded, revealed market liquidity is very low. Like the orange sellers in my example, the short time of quotes makes it very difficult for buyers and sellers to move large volumes without changing the prices.

In their role as liquidity providers, high-frequency traders have taken over the role of market-makers as being traders who do not buy shares to hold in their own right, but simply buy and sell to others and make a profit on this trading. Unfortunately the traditional duty of market-makers to ensure an orderly market, and not to favour themselves over their clients, seems to have been lost in the cracks somewhere.

As Noser points out, there are well-established rules for order book precedence in market-making and there is no obvious reason why high-frequency traders should be exempted from these rules.

As a minimum high-frequency trading needs reforming, with a return to the rules traditionally imposed on market makers, including a minimum required time for a quote to be offered of say five seconds, along with reinstatement of the normal price and time rules for filling orders. (Traditionally market order books are filled first by precedence of price, and then by time of arrival of the quote.)

This would allow competition to revert to that of price and spread, rather than speed. The resultant recreation of meaningful bid-ask spreads, though possibly larger, would be much better at providing signalling of liquidity requirements, which is of course the whole point of market-making in the first place. The increase in price transparency should far outweigh the cost of the free options offered.

Looking more broadly, speed of trading, and narrowness of spread are not the only benefits required from a liquid market. As is seen in Bouchaud et al's paper, high speed does not guarantee the ability to trade a large volume. Similarly, a narrow spread does not mean good value if the upper and lower bands of the spread move against you rapidly as soon as you start trading.

In fact, a good liquid market has a combination of three dimensions, the ability to trade large volumes, at good prices, at high speeds. The way markets are structured allows high-frequency trading to prioritise the advantages of speed at the expense of price and volume.

Supporters of high speed trading show reduced spreads as the main benefit of their technologies, with the implicit assumption that this has clearly reduced costs for all market participants. But the reduced spreads have been accompanied by increased frequencies of trading. It is the belief of the author that the increased speed of trading, and the faster reaction of markets to order flow mean that short term momentum effects have been increased, so obliging all traders to balance their portfolios more frequently.

It is trivially obvious that if spreads are halved, but traders are forced to trade three times more frequently, then overall trading costs have been increased by 50%. If the majority of gains are going to the algorithmic traders, then costs to normal traders have been increased even further. And here 'normal traders' ultimately means the general public as savers, and genuine capitalists raising money to invest in productive capacity.

One possible way to manage this is to change the trading rules so that they also reward providers of volume, longer quotes, and so good stable pricing.

The big advantage in offering larger volume quotes is clearly that more trading can be done faster, and at lower cost. The existence of over-the-counter 'upstairs' markets suggests that institutional investors often want to sell and buy large quantities at the same time, however the ad-hoc nature of upstairs markets can make such exchanges slow and expensive, indeed 'dark-pools' appear to be part of an ongoing process to formalise and automate this upstairs market. Whether 'light-pools' form an extra step in this process remains to be seen.

The big disadvantage of trading large volumes is that it gives a large information signal and causes large adverse movements if only one side of such a potential trade advertises their potential trade.

If more bidders provided longer quotes this would give more quotes available, more price transparency and greater competition. Unfortunately, as discussed above, a long-life quote gives a 'free-option' to traders who can predict the direction the market is going to move. This therefore encourages short quotes, which in a circular reinforcement, encourages rapid price movements.

It is possible that Credit Suisse, or other organisers of 'light pools' may be able to increase the effectiveness and liquidity of their trading platforms if they used rules along the lines of the following for filling orders against the limit order book:

> 1. All quotes to be quoted with both a size and a 'valid-to' time as well as a price. The quote would stand at least to the valid-to time. The valid-to time could be extended, or be rolling from the present time, but the quote could not be cancelled before the valid-to time, and a rolling quote would only be convertible into a valid-to quote of the same length.
> 2. Impose a minimum valid-to time of a few seconds.
> 3. Fill orders firstly according to price.
> 4. Where offers have the same price the offer with the furthest 'valid-to' time is selected first.
> 5. Where offers have the same prices and 'valid-to' time, the offer with the largest volume is selected first.

All incoming orders would follow the same rules, any that crossed the existing order book would be settled immediately, any that don't cross would be obliged to remain on the book until at least the end of the minimum 'valid-to' time.

This would be a 'no-time-wasters' market. It is possible that all quotes submitted would be for the minimum valid-to time, with small quotes competing on price only. However it is the belief of the author that such a market would encourage competition first on length of quote and then on volume.

The minimum time period would form an initial 'level playing field' and would discourage opportunistic bids. Given an existing price level, a new quote on the market that wanted to ensure a sale could simply quote a better price. Alternatively they could put in a quote at the same price but with a later 'valid-to' time. If the extension of time was relatively short, this second course would probably be cheaper than quoting a better price, especially if the market was stable. So at first the market should get a greater amount of quotes going further into the future.

With more bids on each side of the limit book, dealers that had large positions to move would then be able to compete on volume rather than price. If they did this alone in the current HFT market it would be suicidal, but with more 'revealed liquidity' on each side of the book, the proportion of new information revealed would be smaller.

This process should allow more visibility and stability in pricing and so better price discovery. This could then feed back into more competition on quote duration and volume. Ultimately, if this system did work it would have more quotes, more volume and more revealed liquidity than other markets, and ultimately, smaller spreads.

The whole point of the proposed system above is to make traders behave more like fruit stall holders, or better, shop-keepers; to incentivise them to advertise their prices for longer periods and greater amounts of goods, so allowing better competition.

Counter-intuitively, under such a system, much greater liquidity, and better overall price value may also be achieved by limiting the intervals in prices at which shares can be traded and also by limiting the frequencies at which 'valid to' times can finish, say every 2 seconds. Infinite granularity would be reserved just for volume. This would be a reversal of recent history in the management of stock-exchanges. This would prevent price competition at very small fractional levels of price and time, and so encourage more competition on quote time length and volume.

A second area in which the current structure of markets seems sadly lacking is at the opening and closing of sessions. Currently this is commonly done by complex bidding procedures and crossing algorithms to dictate median prices. The suspicion that these procedures don't work; that the median prices are not in fact the market prices, is reinforced by the fact that the majority of trading in equity markets takes place in the first and last hour or so of the trading day. Figure 9.2.1 below gives the price (thick line), and volume (smaller grey shading towards bottom) for shares in HSBC, a large UK bank. Although the scale is a bit small, it can be seen from the volume that the majority of the trading takes place at the start and end of the trading sessions. This is typical of share trading patterns.

Figure 9.2.1 here
[FT.com Markets/data]

The problem here is that as the market opens, liquidity goes from zero to near infinite instantaneously. Conversely, at the close of the market, liquidity goes from infinite to zero instantaneously.

It is well known that increasing liquidity decreases spreads, so conversely, deliberately decreasing liquidity should increase spreads. This suggests an alternative to crossing procedures.

Opening a market could be managed by steadily increasing the liquidity over the first half hour. This could be done easily by opening the market with a very large minimum trade size, in the UK market this would be a minimum multiple of the normal market size 'NMS'. With this large minimum trade size, bids and offers would be a long way apart, and it is very unlikely that any trading would take place. Over the first half hour the minimum bid size would then slowly be moved from a large multiple of NMS to the normal minimum quote size. At some point during this process the bid and offer prices would come close enough for trading to start. This starting point would then be exactly the correct market price. A similar process could be used in reverse for closing markets.

Following the ideas above, it might be better to use the length of time that a quote is held open as the way of manipulating liquidity. At the opening of the market, minimum quote length would be in the order of minutes, and would then be steadily shortened. This would have the same effect of bringing the bid and ask prices together slowly, while having the advantage of not discriminating against small traders.

In fact, although this process would be very useful for restarting a stopped market, it wouldn't generally be necessary. Some commodities markets have already solved this problem. For example the oil futures market run by ICE has trading hours between 01:00am and 11:00pm (UK time). Again the figure below gives price (thick line), and volume (smaller grey shading towards bottom).

Figure 9.2.2 here
[FT.com Markets/data]

Although this might raise fears of traders being forced to work anti-social hours, actually the reverse is true. Trading through the night is low, and then trading and liquidity both rise to a morning peak, followed by a larger afternoon peak before dropping off again. Clearly this has settled to a standard pattern where people who have large trades wait for the liquidity peak to build before they move in to trade.

It would certainly be feasible to do the same for the major stock-exchanges, if only for the larger shares such as those in the FTSE100 index.

All the above are the suggestions of an amateur game theorist. Within economics in recent years there has been an explosion of literature on game theory and auction theory, but this seems to have had little practical input to the trading of financial assets in general and market microstructure in particular. The systematic application of game theory to continuous double auction markets would appear to be a very productive potential future field.

## 10.0  Retail Savings Insurance

The following is a proposal for compulsory default insurance for retail bank deposits. This would not be intended as a realistic way of insuring the deposits, but as a way of introducing market pricing into the risk of government bank deposit insurance. If done correctly this would also reduce the moral hazard element of public assurance of bank deposits.

Realistically, in a democratic capitalist society, a government run central bank will always need to be the lender of last resort and will need to guarantee the deposits of members of the general public to a basic level.

However, such guarantees remove all risk for all but the richest members of the public. It encourages them to move their deposits to the highest interest payers without any need to worry about whether the bank is well run or in danger of collapse.

This then encourages all banks, even the well run, to compete on interest paid while ignoring the risk taken. Indeed the well run banks are forced to match the foolishness of their badly run competitors if they wish to stay in business.

A way to resolve this is to insist that all deposit-taking banks apply compulsory deposit insurance on their deposits. The insurance would be strictly in the form of a percentage charged on the deposits, and this would be displayed in parallel to the interest rate paid by the bank.

It would be illegal for a particular bank to offer its own insurance on its own accounts, and it would be compulsory for banks to offer all alternative insurance from all alternative deposit taking banks.

Bank customers would be able to swap their insurance simply and electronically at any time they wished, from a visible list of alternatives available via the account.

All deposit taking banks would be obliged to offer a price for insurance for all their competitors. They may wish to price their insurance at a high level, but they would be obliged to price, and would be obliged to take on the insurance at the price offered.

In the event of a bank failing, the insuring banks would be obliged to pay the deposits of the insured depositors from their own bank's funds (to avoid spreading systemic risk, reinsurance of this risk would be prohibited; banks would be obliged to carry a portion of funds against these risks on their balance sheets).

The central bank would remain the ultimate insurer of the deposits but would only step in if there was a pattern of systemic risk, and even then only after bank shareholders and all bondholders were wiped out. In the event of a single bank failure due to poor management, the other banks, the insurers, would carry the costs by themselves.

Further rules would apply even in the event of systemic failure. Government deposit guarantee would apply up to a maximum limit (say £100,000), but this maximum guarantee would apply across all deposits for a single person, no matter how many accounts failed at any number of banks. The maximum paid out would be £100,000 even if the person invested £10k in each of 20 different accounts, all of which failed simultaneously. Similarly the government deposit guarantee would only cover £100,000 maximum over any 10-year rolling period.

Individual bank customers would only be able to waive the compulsory bank insurance where they could demonstrate that they already had £100,000 deposited in insured accounts.

Although the above may sound complex, it would be trivial to put in place in a modern electronic retail banking system.

The net effect of this would be to create a market in retail bank deposit insurance. While the Bank of England may have been surprised by the collapse of Northern Rock, Bradford & Bingley and HBOS; the author was not. The rumours of all these impending bank failures were wandering around internet forums from early 2007 onwards. Banking insiders knew that the funding models for these banks were unsustainable and dangerous.

Forcing banks to insure each other's deposits would force banks to price the risk on badly run banks like Northern Rock at higher rates than better run banks such as HSBC and Barclays. By pricing this risk strictly as a percentage rate, the general public would gain direct visibility of the default risk.

Under this regime, a well-run bank might still pay lower interest rates, but would be compensated with even lower insurance rates. This should make the net interest rate; interest less insurance, of the low risk bank better than that of the risky bank. Competition would no longer be on interest rates alone.

With the best will in the world, such a system would not be capable of insuring all deposits in the event of a systemic bubble. But that is not the point.

The point is; that by introducing effective market based pricing of risk, the general public and the banks would be penalised for indulging in the risk-taking that encourages bubbles in the first place.

Additionally, the general rates of insurance should act as both an early warning system for the monetary authorities and even as a counter-cyclical assistance in popping bubbles in the first place.

In normal times, insurance rates for all but the most foolish of banks should be ridiculously low. In the event of the economy moving into bubble conditions, insurance rates would start to creep up on the riskiest banks. This would then start to pass on the infection, via the insurance, to other banks, but at a much earlier stage than normally happens when entering a financial crisis. Faced with the obligation of holding more reserves on their balance sheets to cover the deposit failure of others, all banks would be obliged to cut back on credit in general. All banks would be affected, but with the strongest effects on the worst run and most highly leveraged banks.

Monitoring of individual and overall insurance rates would give the central banks live data on the perceived risks of the banks in their charge, as well as the financial system as a whole.

## 11. Conclusions

This paper does not in fact have strong definite conclusions. The paper was intended more to be a starting point than a conclusion. All the ideas above are speculative. It remains to be demonstrate that companies do in fact operate as market-makers and that liquidity is a state variable. Similarly, while the proposals for reducing chaotic behaviour in housing seem sensible, those proposed for managing stock markets may not be practical.

The main aim of this paper has been to stimulate a new way of thinking about markets in economics and finance. This is a way of thinking that is strongly predicated on the view that markets are inherently dynamic.

If this paper helps to stimulate new research and experimentation based on a dynamic basis it will have achieved its aim.

## 12. Acknowledgements

I would also like to give thanks to Patricia Chelley-Steely for giving me important insights into the role of market-microstructure in general and liquidity in particular.

## 16. Figures

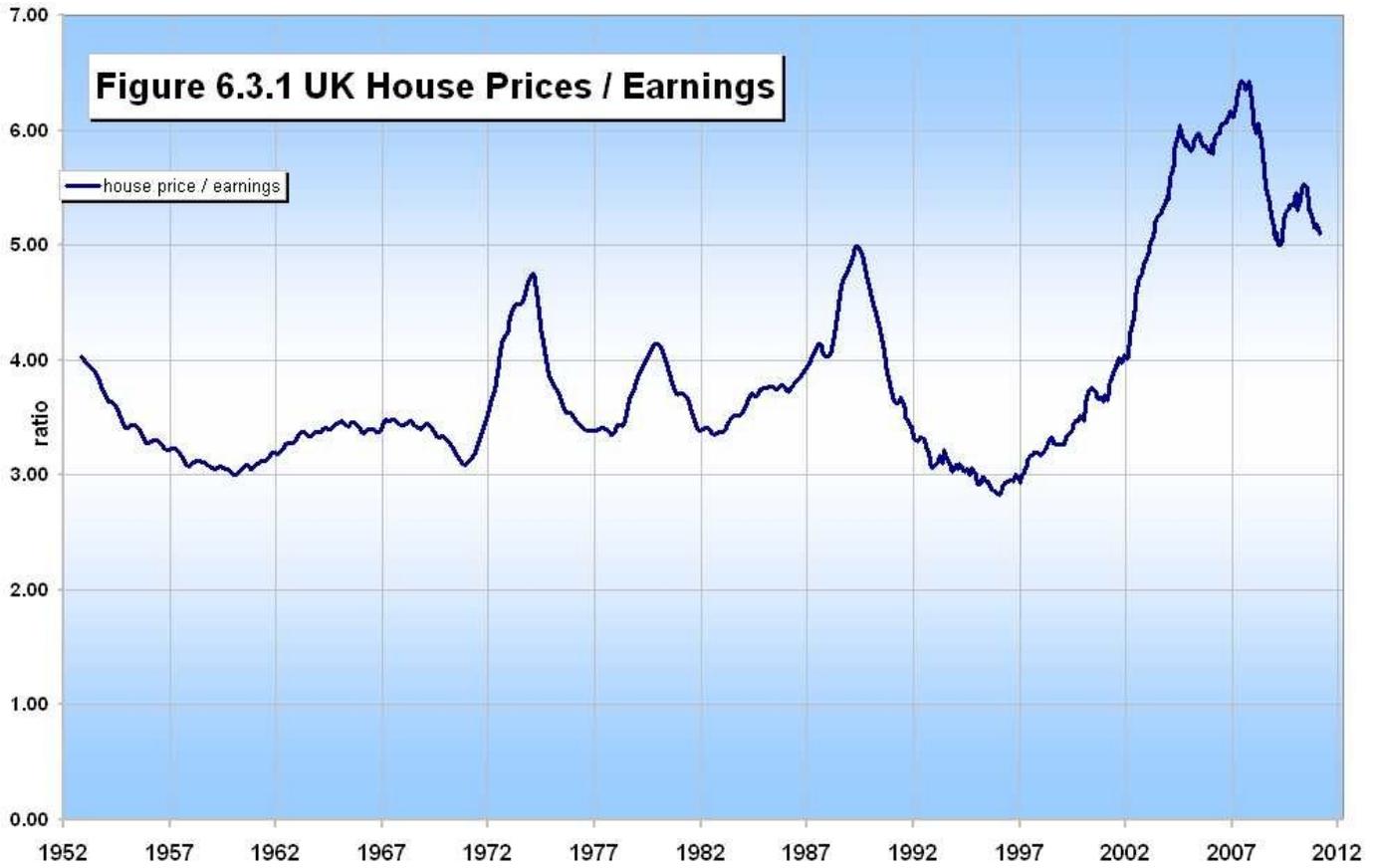

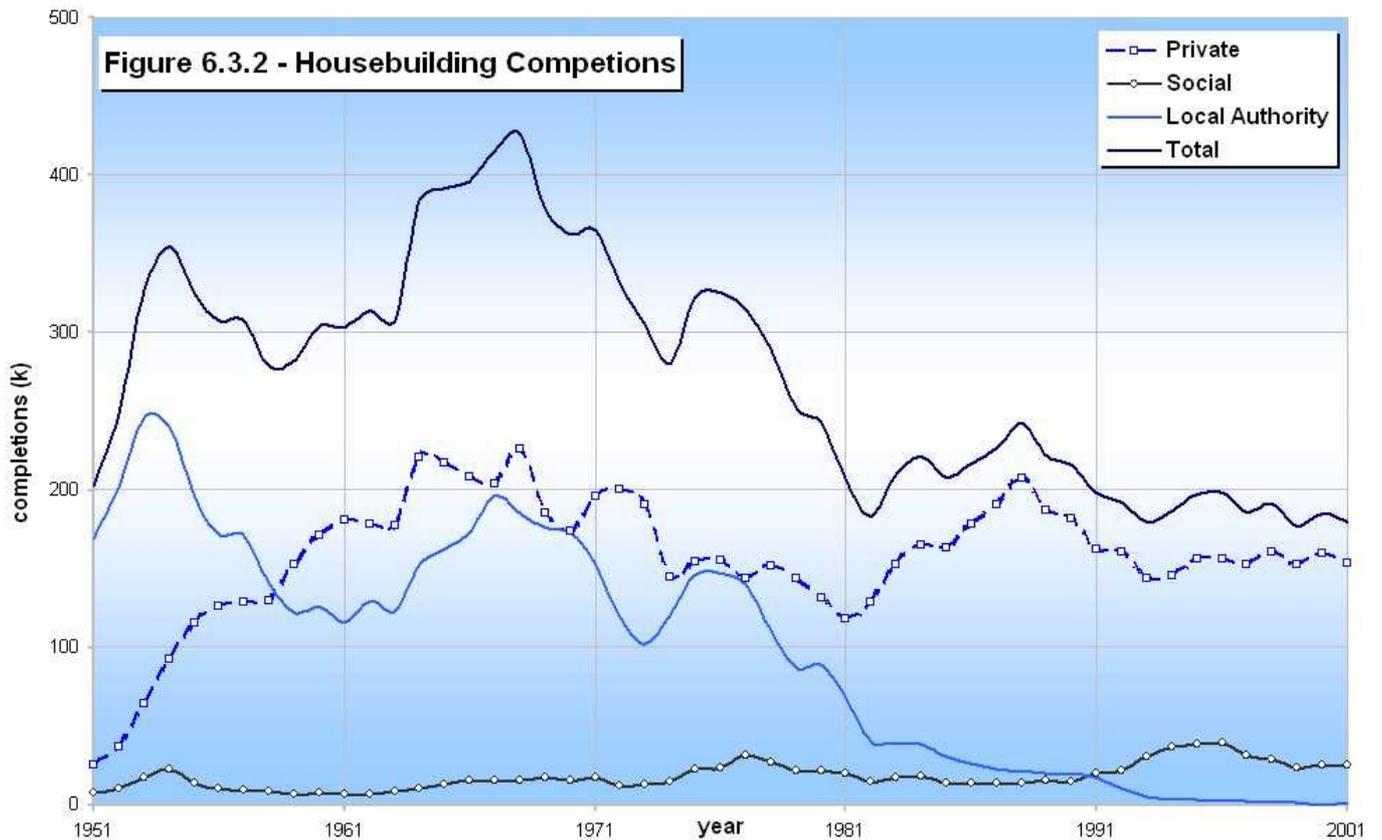

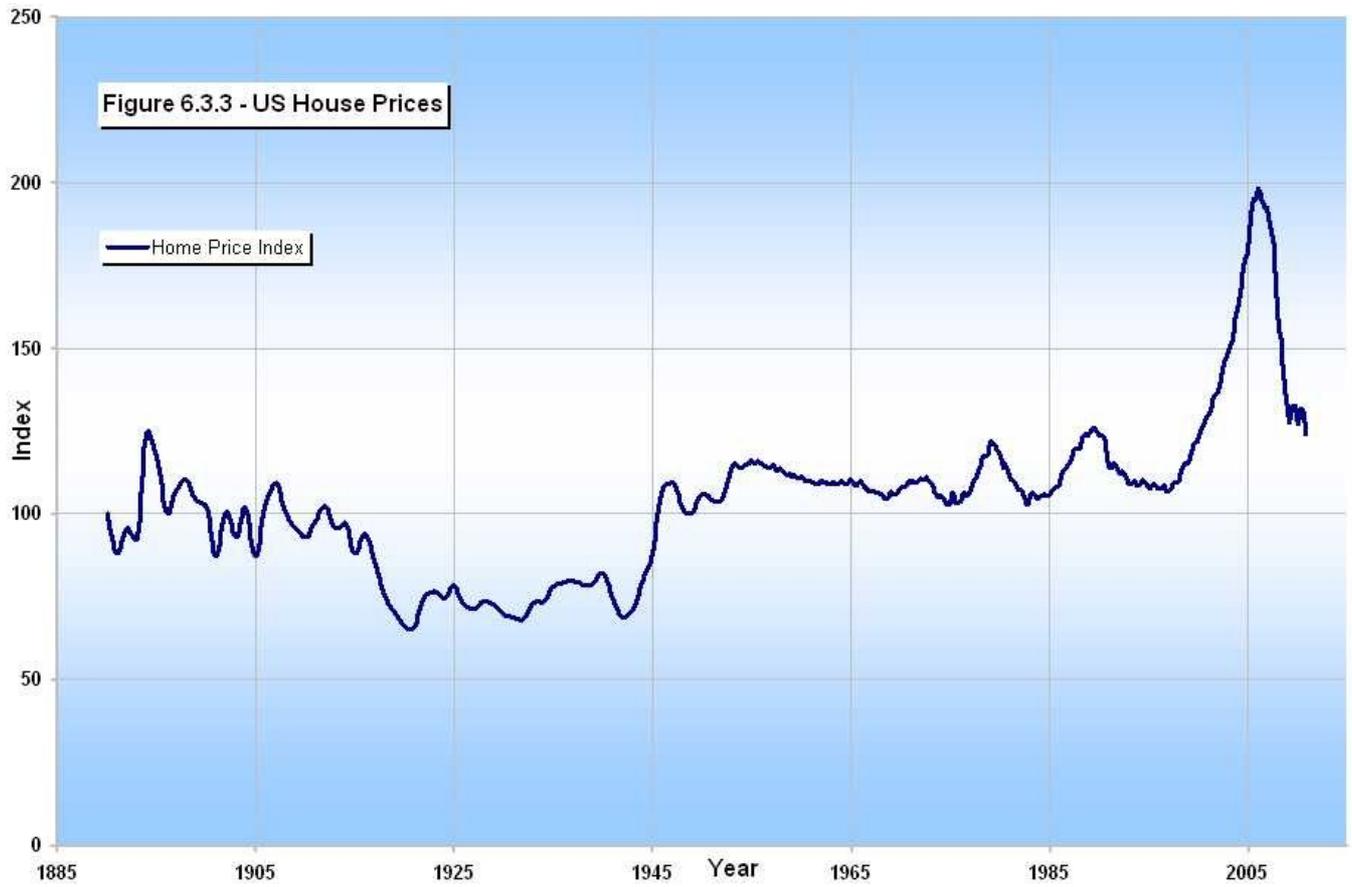

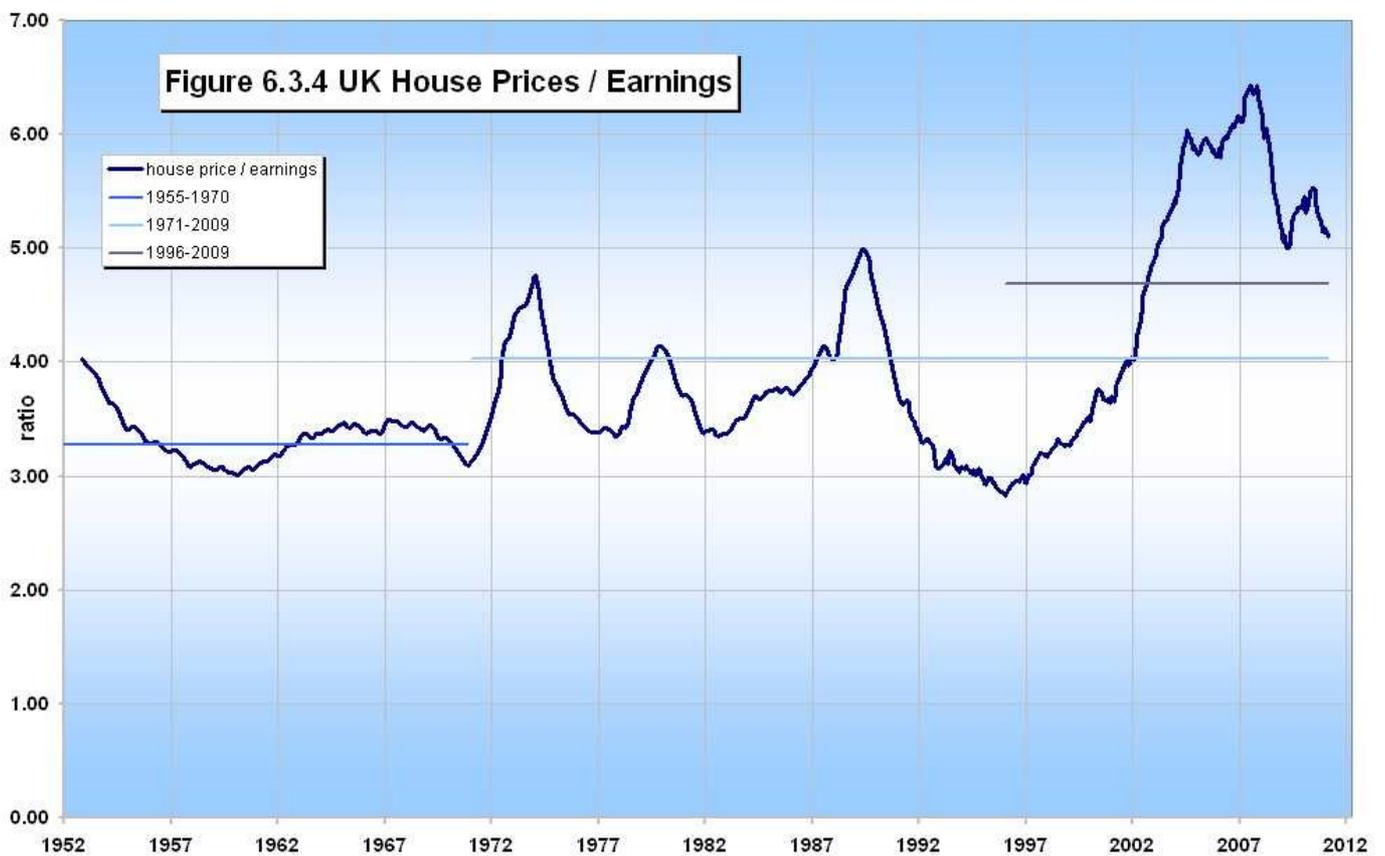

| Fixed-rate vs. floating rate systems | |
|---|---|
| Land | Rate adjustment (percentage of new business) * |
| Belgium | F (75%), M (19%), V (6%) |
| Denmark | F (75%), M (10%), V (15%) |
| Germany | Mainly F and M |
| Greece | F (5%), M (15%), V (80%) |
| Spain | V (more than 75%) |
| France | F/M/O (86%), V (14%) |
| Ireland | V (70%), otherwise mainly M |
| Italy | F (28%) |
| Luxembourg | V (90%) |
| Netherlands | F (74%), M (19%), V (7%) |
| Austria | F (75%), V (25%) |
| Portugal | Mainly V |
| Finland | F (2%), V 97%), O (1%) |
| United Kingdom | V (72%), M (28%) |

\* Fixed (F): interest rate fixed for more than five years or until final maturity;
Mixed (M): interest rate fixed for one to five years;
Variable (V): interest rate renegotiable after one year or tied to market rates
or adjustment at the lender's discretion
Other (O)

Source: ECB (2003)

Figure 6.3.5 [Hess & Holzhausen]

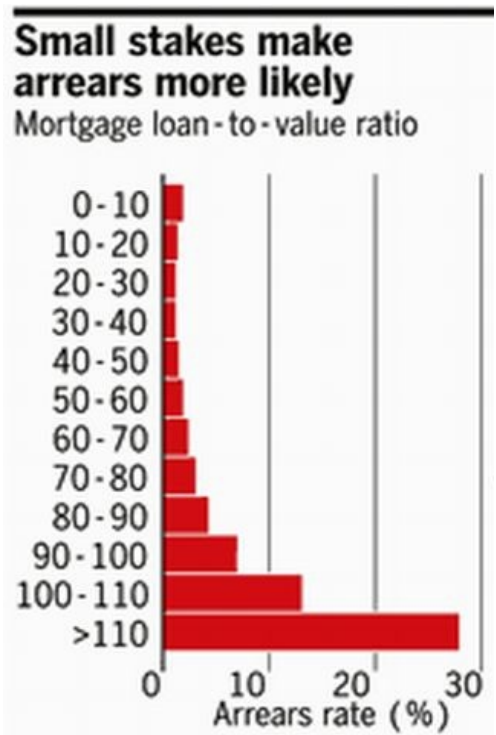

Figure 6.3.6

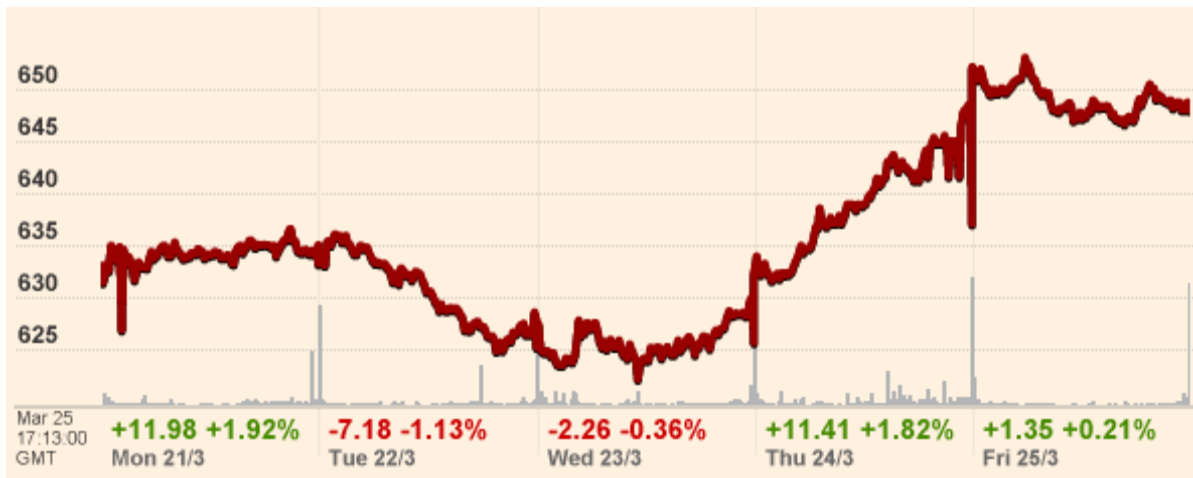
Figure 9.2.1

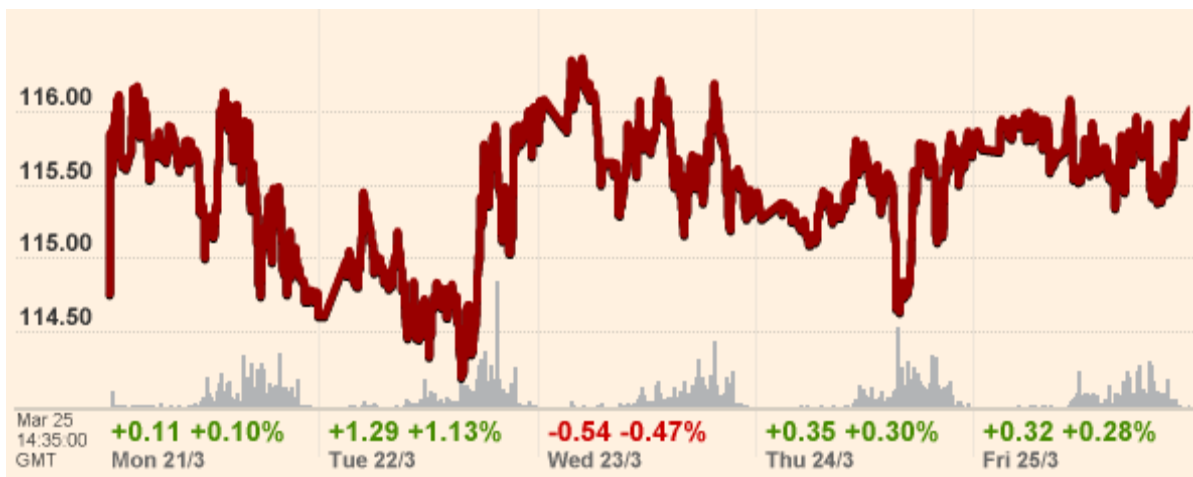
Figure 9.2.2